# Topolectrical-circuit octupole insulator with topologically protected corner states

Jiacheng Bao[1*], Deyuan Zou[2*], Weixuan Zhang[2], Wenjing He[1], Houjun Sun[1$], and Xiangdong Zhang[2+]

[1] Beijing Key Laboratory of Millimeter wave and Terahertz Techniques, School of Information and Electronics, Beijing Institute of Technology, Beijing 100081, China

[2] Key Laboratory of advanced optoelectronic quantum architecture and measurements of Ministry of Education, Beijing Key Laboratory of Nanophotonics& Ultrafine Optoelectronic Systems, School of Physics, Beijing Institute of Technology, 100081, Beijing, China

*These authors contributed equally to this work. $+Author to whom any correspondence should be addressed. E-mail: sunhoujun@bit.edu.cn；zhangxd@bit.edu.cn

## Abstract

Recent theoretical studies have extended the Berry phase framework to account for higher electric multipole moments, quadrupole and octupole topological phases have been proposed. Although the two-dimensional quantized quadrupole insulators have been demonstrated experimentally, octupole topological phases have not previously been observed experimentally. Here we report on the experimental realization of classical analog of octupole topological insulator in the electric circuit system. Three-dimensional topolectrical circuits for realizing such topological phases are constructed experimentally. We observe octupole topological states protected by the topology of the bulk, which are localized at the corners. Our results provide conclusive evidence of a form of robustness against disorder and deformation, which is characteristic of octupole topological insulators. Our study opens a new route toward higher-order topological phenomena in three-dimensions and paves the way for employing topolectrical circuitry to study complex topological phenomena.

Topological phases exhibit some most striking phenomena in modern physics. A prominent feature of a topological phase is the emergence of topologically protected edge states, which are robust against local perturbations and play a crucial role in the topological functionality of the underlying system [1-4]. Topological insulators as important platforms for realizing topologically protected boundary states have attracted great attention in recent years [2-5]. They have been constructed in various systems ranging from traditional electronic setups [2-5] to mechanical [6], electromagnetic [7-11] and acoustic [12-14] structures governed by classical wave equations.

Recently, a new class of symmetry-protected topological insulators, higher-order topological insulator, has drawn research interest [15-17]. Unlike conventional first-order topological insulators, two-dimensional (2D) second-order topological insulators have topologically protected corner states, and corresponding three-dimensional (3D) systems have topological gapless modes on the hinges. In some crystalline structures, the topological corner and hinge states can arise only from the nontrivial bulk topology when the lattice termination is compatible with the crystal symmetries. The first prediction of a second-order topological insulator, based on quantized quadrupole polarization, was demonstrated in classical mechanical [18] and microwave [19] systems, as well as in electrical circuits [20, 21]. The other kinds of second- and third-order topological phases have also been observed in acoustic, photonic and electronic systems [22-44]. However, the experimental realization of much higher topological insulators, such as octupole topological insulators, remains challenging.

Here we report on the experimental realization of classical analog of octupole topological insulator using 3D topolectrical circuits. The circuits consist of some basic circuit elements: capacitors and inductors, which are constructed experimentally. The octupole topological phases, which are localized at the corners, are observed. The octupole topological insulators that depend only on the bulk topology are demonstrated experimentally.

***Theoretical design of topolectrical-circuit octupole insulator.*** ─LC circuit models are well suited to the implementation of tight-binding models, because it is possible to establish a correspondence between individual hopping terms in the tight-binding model and individual components in its circuit realization [45-51]. This potential has already been exploited experimentally to implement conventional and quadrupole topological insulators [20, 21]. Now, we extend the LC circuit model to study three-dimensional octupole insulator. We consider a 3D circuit unit cell shown in Fig.1a, which a tight-binding model with quantized octupole insulator can be constructed by using it.

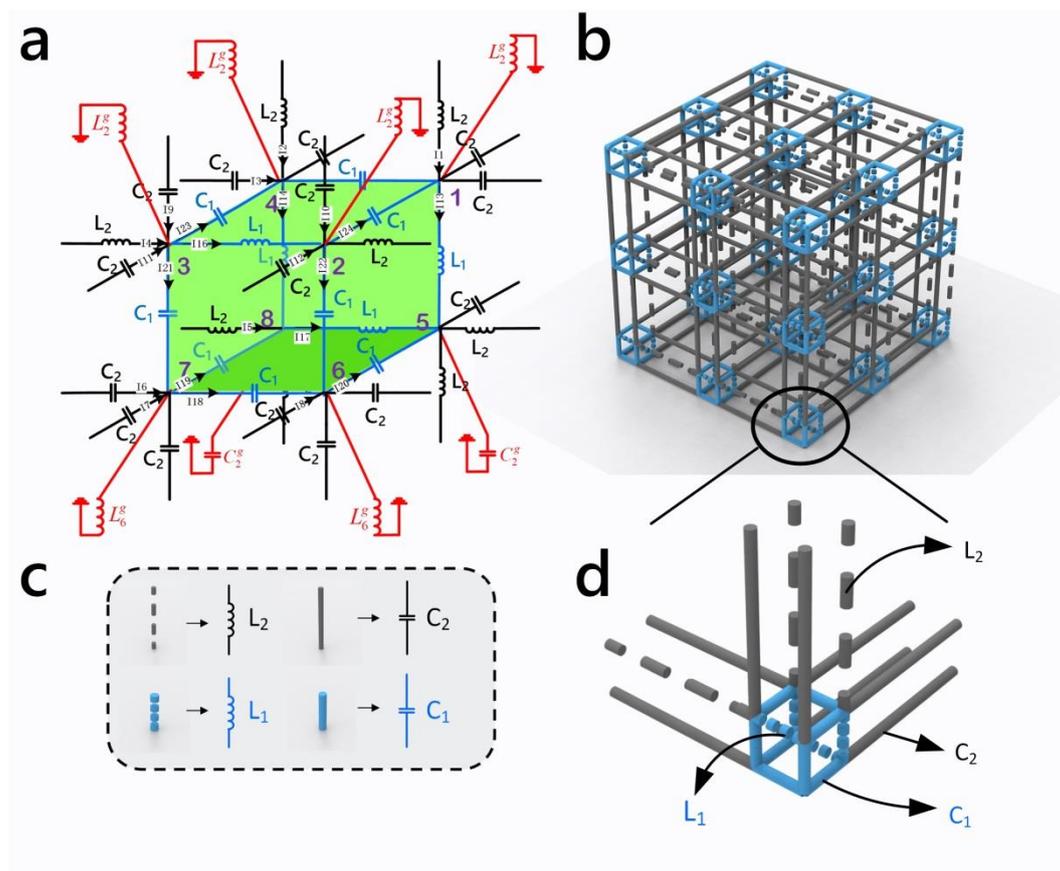

**Fig.1 Circuit model of the octupole topological insulator. a, Unit cell of the circuit. The capacitors and inductors are marked in the figure. Blue and black circuit elements correspond to weak and strong bonds in a tight-binding or mechanical analogue of the circuit. Red circuit elements connect to the ground for the sites 1, 2, 3, 4, 5, 6, 7 and 8 via inductivities and capacitors. b, Circuit scheme with 3x3x3 unit cells. c, the corresponding marks for inductivities and capacitors in Fig.2b. d, Enlarged a corner region including a unit cell in Fig.1b.**

The unit cell consists of two pairs of capacitors and inductors, $(L_1, C_1)$ and $(L_2, C_2)$, which satisfy the relations $C_2 = \lambda C_1$ and $L_1 = \lambda L_2$. So that they have a same resonance frequency $\omega_0 = (L_1 C_1)^{-1/2} = (L_2 C_2)^{-1/2}$. Here $\lambda$ is real positive parameter. The red parts are the groundings in eight sites. It is worthy to note that different values of grounded capacitors/inductances should be used to make the onsite potential on each lattice site become zero at the working frequency (see details in Ref. [52]). So the sites 1, 2, 3 and 4 are connected to the ground via an inductivity $L_2^g = L_1/(1+\lambda)$, the sites 6 and 7 are connected to the ground via an inductivity $L_6^g = L_1/(3+3\lambda)$, the sites 5 and 8 are connected to the ground via an capacitor $C_2^g = C_1 + C_2$. So, we can use the parameters $\omega_0$ and $\lambda$ to describe the topological properties of the circuit unit cell.

In order to realize an octupole insulator with topologically protected corner states, the system should have three anti-commuting mirror symmetries [15]. So, the same anti-commuting mirror symmetries should also be owned by our classical circuit systems. According to this condition, we build a classical analogue of the electric octupole insulator, which has three anti-commuting mirror symmetries ($M_x$, $M_y$ and $M_z$) for modes near a specific frequency $\omega_0$. We first consider an infinite periodic structure realized by the above circuit unit cell, and study the bulk properties of the system, before analyzing boundary modes. We then analyze the circuit with periodic boundary conditions in momentum space. In circuit systems, we can derive the circuit Laplacian by the Kirchhoff's current law [45-51],

$$J(\omega) = i\omega C - \frac{i}{\omega} W, \qquad (1)$$

where $J(\omega)$ is called as the circuit Laplacian, and $C$ is capacitance, $W$ is the inverse inductivity $W = L^{-1}$. The off-diagonal components of matrix $C$ are the capacitances $C_{ab}$ between nodes a and b. The diagonal components of matrix $C$ are total node capacitances $C_{aa}$. The matrix $W$ is the same to $C$. Under the condition $\omega = \omega_0$, one of the bonds on each square in the green cube shown in Fig.1a has a negative sign, and each square plaque contains $\pi$-flux in each direction. Using Fourier transform, the circuit Laplacian $J(\omega)$ in Eq.(1) can be transformed to the following $J_\lambda(\omega_0, q)$ in the momentum space [52]:

$$J_\lambda(\omega_0, q) = i\sqrt{\frac{C_1}{L_1}}[\lambda \sin q_y \Gamma'_1 + (1+\lambda \cos q_y)\Gamma'_2$$
$$+ \lambda \sin q_x \Gamma'_3 + (1+\lambda \cos q_x)\Gamma'_4$$
$$+ \lambda \sin q_z \Gamma'_5 + (1+\lambda \cos q_z)\Gamma'_6], \quad (2)$$

where $\lambda = C_2/C_1$, $q_i (i=x,y,z)$ is the phase of Block wave vector propagating along the x, y and z directions, respectively. $\Gamma'_i = \sigma_3 \otimes \Gamma_i$ for i=0,1,2,3,4, and $\Gamma'_5 = \sigma_2 \otimes I_{4\times 4}$ $\Gamma'_6 = i\Gamma'_0\Gamma'_1\Gamma'_2\Gamma'_3\Gamma'_4\Gamma'_5$. Here, $\Gamma_0 = \tau_3\sigma_0$, $\Gamma_k = -\tau_2\sigma_k$, and $\Gamma_4 = \tau_1\sigma_0$, for k = 1, 2, and 3. This circuit Laplacian satisfies

$$M_x J_\lambda(\omega_0, q_x, q_y, q_z) M_x^{-1} = J_\lambda(\omega_0, -q_x, q_y, q_z)$$
$$M_y J_\lambda(\omega_0, q_x, q_y, q_z) M_y^{-1} = J_\lambda(\omega_0, q_x, -q_y, q_z), \quad (3)$$
$$M_z J_\lambda(\omega_0, q_x, q_y, q_z) M_z^{-1} = J_\lambda(\omega_0, q_x, q_y, -q_z)$$

where $M_x = \sigma_0 \otimes \sigma_1 \otimes \sigma_3$, $M_y = \sigma_0 \otimes \sigma_1 \otimes \sigma_1$ and $M_z = \sigma_1 \otimes \sigma_3 \otimes \sigma_0$ represent reflection symmetry operators, which obey $\{M_i, M_j\} = 0$ ($i,j=x,y,z$ and $i\neq j$). Here $\tau_i(i=1,2,3)$ and $\sigma_i(i=1,2,3)$ are Pauli matrices, and $\tau_0$, $\sigma_0$ are the $2\times 2$ identity matrices. It is found that besides an overall factor of i, $J_\lambda(\omega_0, q)$ in our circuit takes exactly the same form as the Bloch Hamiltonian matrix of the octupole insulator introduced in ref. 14. So that if $\lambda \neq 1$ the spectrum of $J_\lambda(\omega_0, q)$ is gapped. When $\lambda > 1$, the circuit is an octupole circuit. If $\lambda < 1$, the circuit is a trivial circuit.

Now, we turn to a circuit with open boundary conditions as shown in Fig.1b to realize topologically protected corner modes. To achieve this, two conditions must be satisfied. The first one is that the symmetries protecting the topological feature can not be broken by the boundary. The second one is that the boundary should not cut through the unit cell. To satisfy the first condition, we let the diagonal elements of $J_\lambda(\omega)$ to vanish at the resonance frequency $\omega_0$ to protect the symmetry of the circuit. The diagonal elements of $J_\lambda(\omega)$ are the circuit Laplacian in each site including bulk, surface, edge and corner. So, we fix the circuit elements (capacitor and/or inductor) that connect each site to the ground to meet this condition [52]. For the second condition, we let every corner to end at a unit cell to meet it.

With all of the conditions and theories discussed above, we finally construct a topolectrical-circuit octupole insulator shown in Fig.1b. We terminate each edge of the circuit with a unit cell. So, the circuit satisfies all the symmetries $M_x$, $M_y$ and $M_z$, and topological corner modes could thus be protected at each corner. To prove the validity of our circuit, we calculate the spectrum of circuit Laplacian as a function of the driving frequency. In the calculation, the parameters $C_1$ and $C_2$ are taken as 1nF and 3.3nF, respectively, and $L_1$ and $L_2$ are taken as 3.3μH and 1μH. The results are shown in Fig.2a, where an isolated mode clearly crosses the gap. It represents the zero-energy eigenvalue of $J(\omega)$ at $\omega = \omega_0$, which corresponds to the topological corner mode.

Next, we calculate the expected frequency difference between the impedances of the bulk, edge, surface and corner modes. The most natural measurement on a circuit is the impedance response $Z_{ab}(\omega)$, which is the ratio of the voltage between two nodes a and b due to a current. Mathematically, $Z_{ab}(\omega)$ can be expressed as

$$Z_{ab}(\omega) = G_{aa}(\omega) + G_{bb}(\omega) - G_{ab}(\omega) - G_{ba}(\omega)$$
$$= \sum_n \frac{|\phi_n(a) - \phi_n(b)|^2}{j_n(\omega)} \quad , \quad (5)$$

where $G_{ab}(\omega) = J^{-1}_{ab}(\omega)$ is the circuit Green's function, and $j_n(\omega)$ is the eigenvalue of $J_{ab}(\omega)$, which satisfies $J_{ab}(\omega) = \sum_n j_n(\omega) |\phi_n(a)\rangle\langle\phi_n(b)|$. We can see that the impedance can be determined by the smallest eigenvalue $j_n(\omega)$. From the spectrum of circuit Laplacian, we can find that there exist a gap and zero eigenvalue $j_n(\omega) = 0$ in the corner as $\omega = \omega_0$. This phenomenon leads to a large impedance for the corner mode, but not for the other modes. The analysis is agreement with the simulation of the impedance for the circuit. Fig.2b shows the simulated result of impedances as a function of the driving frequency. It is seen clearly that the impedance is extremely high for the corner mode as $\omega = \omega_0$, which is different from the low impedances for the bulk, edge and surface modes. The distributions of zero-energy eigenvalues $j_n(\omega) = 0$ at $\omega = \omega_0$ which correspond to the corner mode are also plotted in

Fig.2c. It is shown clearly that strong zero-energy eigenvalues appear at eight corner positions.

The above discussions only focus on the case with a sample size. In fact, we have also calculated impedance responses for the samples of other sizes. With the increase of sample size, similar phenomena can be found. In addition, we have also studied the effect of random capacitors and inductors in bulk of the sample on the corner modes. Our calculated results show that strong corner modes always appear when the capacitors and inductors vary randomly within the sample or on the surface (except at the corner positions). The most important property of the topologically protected corner state is that it is robust against disorders. We note that the corner state always exist even the value of used capacitors and inductors on the different positions (except for corners) are randomly varied (the relationships of $C_2/C_1 > 1$ and $L_1/L_2 > 1$ should be satisfied), which is similar to the case in Ref. [15]. In the following, we provide the corresponding experimental results.

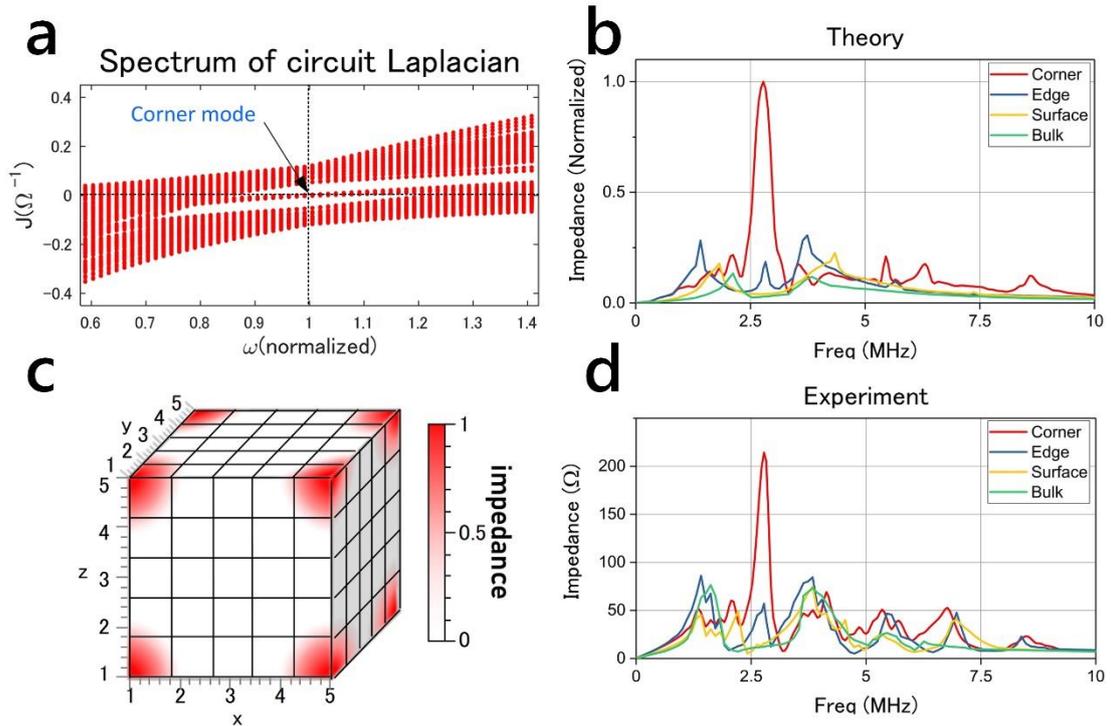

**Fig.2 Comparison of experimental and theoretical results for the circuit spectrum and corner mode. a,** Theoretical spectrum of the circuit Laplacian $J(\omega)$ as a function of the driving frequency. All frequency scales are normalized to the resonance frequency $\omega_0$. An isolated mode crossing the gap, which corresponds to a zero-energy eigenvalue of $J(\omega)$ at $\omega = \omega_0$, is clearly visible. It corresponds to the topological corner mode. The calculation includes a random disorder of 1% for all capacitors and 2% for all inductors. The values of $C_1$ and $C_2$ are taken as 1nF and 3.3nF, and $L_1$ and $L_2$ are taken as 3.3μH and 1μH. b, Theoretical results of the impedance between two nearest-neighbour sites at the corner, edge, surface and in the bulk. c, The distributions of zero-energy eigenvalues for the sample with 3x3x3 unit cells. d, Experimental results of the impedance corresponding to b.

*Experimental observation of octupole topological phases.* 一In order to observe octupole topological phases experimentally, the circuit with $3.0\times 3.0\times 3.0$ unit cells, corresponding to the above theoretical scheme, is designed. Image of the experiment sample is shown in Fig.3a. For the convenience of experiment, we cut the total cube in Fig.1b into six sides. We make these sides on three printed circuit boards (PCB) in order. Capacitors and inductors connect adjacent sides on every site. The different connection ways result in the difference of groundings. Thus, the six sides are divided into two types: A and B. The images of two different sides A and B are shown in Fig.3c and 3d, respectively. The detailed description is given in Ref. [52]. The fabricated sample has exactly the same construction shown in Fig.1b. This can be seen more clearly from Fig.3b, which shows enlarged image at a corner of the PCB. The blue and black parts in Fig.3b represent the capacitors and inductors on one side. The red parts are the capacitors and inductors connecting adjacent sides. The yellow parts are the capacitors and inductors connecting each site to ground. The line-via hole is used to pass wire which connects adjacent sides on different PCB. The Smp connector being located at each site is used for measurement. The copper pillar is connected to ground, which could also support three PCBs.

As for the design of PCB, all PCB traces have a relatively large width (0.7mm) to reduce the parasitic inductance of the traces. The spacing between lines is large enough to avoid spurious inductive coupling. The parameters of circuit elements are taken as identical with the above theoretical calculations. The tolerance of the circuit elements is 1% which can avoid the experiment error. We set $\lambda = C_2/C_1 = L_1/L_2$ to be 3.3, and the resonance frequency to be 2.77 MHz. We use WAYNE KERR precision impedance analyzer to measure the impedance of the circuit as a function of the driving frequency. The experimental results are shown in Fig.2d. The excellent agreement between experimental results and the theoretical predictions have been observed. The theoretical impedance corner peak is normalized to unity, and the experimental impedance corner peak reaches 215 Ω. The red line represents the corner mode, which is extremely high in the resonance frequency. The other lines, which are blue, yellow and green lines, have a small impedance in resonance frequency. The experimental results for the impedance distributions at the frequency $\omega_0$ are also provided in Fig.3c and 3d for sides A and B, respectively. The red parts in Fig.3c mean the appearance of large impedance in the corner. We can see that only side A have red parts but not B, which means the large impedance in eight corners, but not in edges, surfaces and bulk. This means that our experiment successfully demonstrates the existence of corner state in the octupole topological circuit.

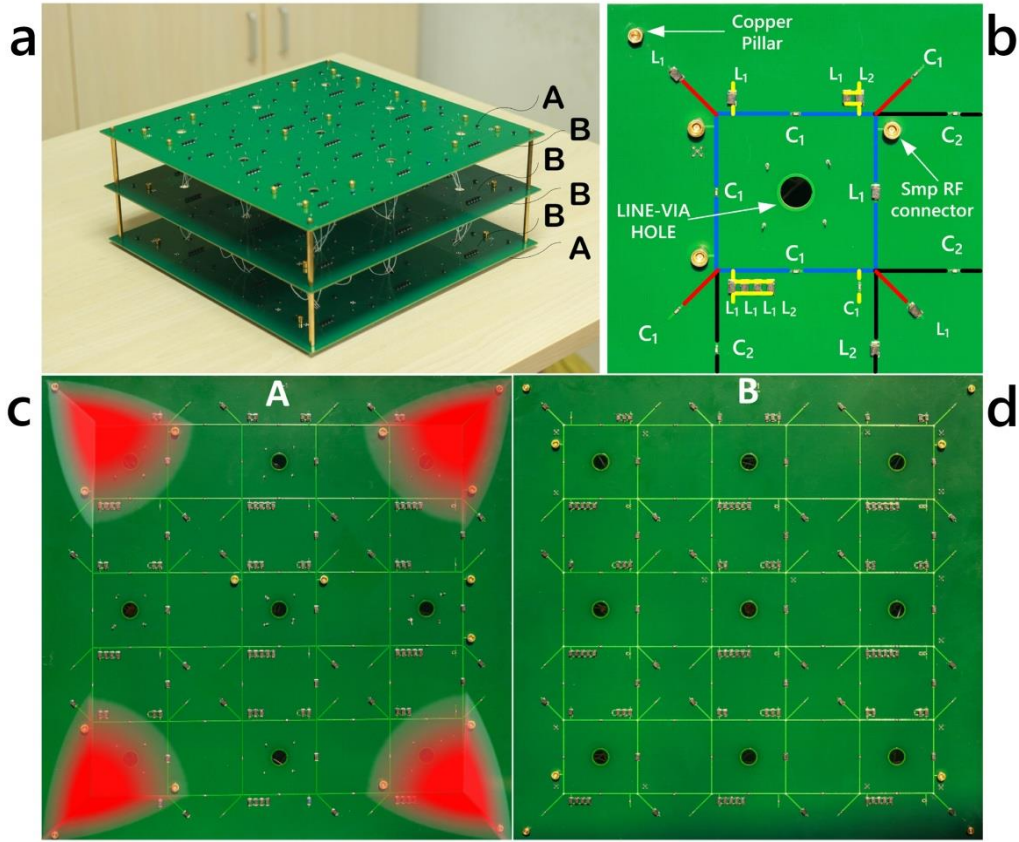

**Fig.3 Experimental sample of the electrical circuit exhibiting topological corner states. a, Image of the printed circuit board of the octupole topological insulator with three unit cells, corresponding to Fig.1(b). b, Enlarged a corner region including a unit cell in Fig.3a. c and d, show the images of the printed circuit board for two different sides A and B, respectively. The impedance distributions at the frequency $\omega_0$ are also shown.**

In summary, we have provided experimental evidence of octupole topological phases of matter. Our circuit implementation of the octupole topological insulator has confirmed the existence of the theoretically predicted corner modes and firmly established their origin from the bulk octupole topology. Before this work, only quadrupole topological insulator in 2D has been realized experimentally in many classical systems, such as photonic, phononic and electronic circuit systems [15-21]. However, the octupole topological insulator in 3D has never been realized due to the difficulty in structural design and sample fabrication. In this work, we firstly construct the 3D octupole topological-circuit, and the associated 0D corner state has also been observed. Although the extension of tight-binding lattice modal from 2D to 3D seems direct, the actual circuit design is not straightforward, e.g. the value of grounded capacitor/inductance on different locations (corner, edge, surface and bulk) should be designed subtly. In this case, our study opens a new route toward higher-order topological phenomena in high dimensions and paves the way for employing topolectrical-circuit to study complex topological phenomena, offers possibilities to control electrical signals in unprecedented ways.

This work was supported by the National key R & D Program of China under Grant No. 2017YFA0303800 and the National Natural Science Foundation of China (No.91850205 and No.61421001).

# Supplementary Information for

**Topolectrical-circuit octupole insulator with topologically protected corner states**

Jiacheng Bao[1*], Deyuan Zou[2*], Weixuan Zhang[2], Wenjing He[1], Houjun Sun[1$], and Xiangdong Zhang[2+]

[1] Beijing Key Laboratory of Millimeter wave and Terahertz Techniques, School of Information and Electronics, Beijing Institute of Technology, Beijing 100081, China

[2] Key Laboratory of advanced optoelectronic quantum architecture and measurements of Ministry of Education, Beijing Key Laboratory of Nanophotonics& Ultrafine Optoelectronic Systems, School of Physics, Beijing Institute of Technology, 100081, Beijing, China

*These authors contributed equally to this work.$+Author to whom any correspondence should be addressed. E-mail: sunhoujun@bit.edu.cn；zhangxd@bit.edu.cn

## S1. Derivation of the circuit Laplacian of the infinite 3D octupole topolectrical circuit in momentum space.

We consider the model shown in Fig.1a. There are voltages $V_1$-$V_8$ at the eight sites and currents $I_1$-$I_{24}$ at the circuit branches. Using the Kirchhoff's current formula, we have

$$I_1 + I_{15} + I_{10} = I_{13} + I_3 e^{iq_x} + I_{12} e^{iq_y} + G, \quad (1)$$

$$I_{11} + I_4 + I_9 = I_{23} + I_{21} + I_{16} + G, \quad (2)$$

$$I_2 + I_3 + I_{23} = I_{15} + I_{14} + I_{11} e^{iq_y} + G, \quad (3)$$

$$I_{16} + I_{12} + I_{10} = I_{22} + I_4 e^{iq_x} + I_{24} + G, \quad (4)$$

$$I_{17} + I_{20} + I_{13} = I_8 e^{iq_y} + I_5 e^{iq_x} + I_1 e^{iq_z} + G, \quad (5)$$

$$I_7 + I_{21} + I_6 = I_{18} + I_{15} + I_9 e^{iq_z} + G, \quad (6)$$

$$I_{19} + I_5 + I_{14} = I_7 e^{iq_y} + I_{17} + I_2 e^{iq_z} + G, \quad (7)$$

$$I_8 + I_{18} + I_{22} = I_{20} + I_6 e^{iq_x} + I_{10} e^{iq_z} + G, \quad (8)$$

where G is grounding part, $q_i\ (i = x, y, z)$ represent the current direction. So we can write the currents that flow into each site as,

$$I_a = \frac{-3V_a}{i\omega L_1} + \frac{-3V_a}{i\omega L_2} + i\omega C_1(V_c - V_a) + i\omega C_1(V_d - V_a) + i\omega C_2(V_e e^{-iq_z} - V_a)$$
$$-i\omega C_1(V_a - V_e) - i\omega C_2(V_a - V_c e^{iq_x}) - i\omega C_2(V_a - V_d e^{iq_y})$$
(9)

$$I_b = \frac{-V_b}{i\omega L_1} + \frac{-V_b}{i\omega L_2} + i\omega C_2(V_d e^{-iq_x} - V_b) + i\omega C_2(V_f e^{-iq_z} - V_b) + \frac{(V_c e^{-iq_y} - V_b)}{i\omega L_2}$$
$$-i\omega C_1(V_b - V_d) - i\omega C_1(V_b - V_f) - \frac{(V_b - V_c)}{i\omega L_1}$$
(10)

$$I_c = \frac{-V_c}{i\omega L_1} + \frac{-V_c}{i\omega L_2} + i\omega C_2(V_a e^{-iq_x} - V_c) + i\omega C_2(V_g e^{-iq_z} - V_c) + \frac{(V_b - V_c)}{i\omega L_1}$$
$$-i\omega C_1(V_c - V_a) - i\omega C_1(V_c - V_g) - \frac{(V_c - V_b e^{iq_y})}{i\omega L_2}$$
(11)

$$I_d = \frac{-3V_d}{i\omega L_1} + \frac{-3V_d}{i\omega L_2} + i\omega C_1(V_b - V_d) + i\omega C_2(V_a e^{-iq_y} - V_d) + i\omega C_2(V_h e^{-iq_z} - V_d)$$
$$-i\omega C_1(V_d - V_a) - i\omega C_1(V_d - V_h) - i\omega C_2(V_d - V_b e^{iq_x})$$
(12)

$$I_e = -V_e i\omega C_1 - V_e i\omega C_2 + \frac{(V_g - V_e)}{i\omega L_1} + \frac{(V_h - V_e)}{i\omega L_1} + i\omega C_1(V_a - V_e)$$
$$-\frac{(V_e - V_g e^{iq_x})}{i\omega L_2} - \frac{(V_e - V_h e^{iq_y})}{i\omega L_2} - i\omega C_2(V_e - V_a e^{iq_z})$$
(13)

$$I_f = \frac{-V_f}{i\omega L_1} + \frac{-V_f}{i\omega L_2} + \frac{(V_h e^{-iq_x} - V_f)}{i\omega L_2} + i\omega C_1(V_b - V_f) + i\omega C_2(V_g e^{-iq_y} - V_f)$$
$$-\frac{(V_f - V_h)}{i\omega L_1} - i\omega C_2(V_f - V_b e^{iq_z}) - i\omega C_1(V_f - V_g)$$
(14)

$$I_g = \frac{-V_g}{i\omega L_1} + \frac{-V_g}{i\omega L_2} + \frac{(V_e e^{-iq_x} - V_g)}{i\omega L_2} + i\omega C_1(V_f - V_g) + i\omega C_1(V_c - V_g)$$
$$-\frac{(V_g - V_e)}{i\omega L_1} - i\omega C_2(V_g - V_f e^{iq_y}) - i\omega C_2(V_g - V_c e^{iq_z})$$
(15)

$$I_h = -V_h i\omega C_1 - V_h i\omega C_2 + \frac{(V_e e^{-iq_y} - V_h)}{i\omega L_2} + i\omega C_1(V_d - V_h) + \frac{(V_f - V_h)}{i\omega L_1}$$
$$-\frac{(V_h - V_e)}{i\omega L_1} - \frac{(V_h - V_f e^{iq_x})}{i\omega L_2} - i\omega C_2(V_h - V_d e^{iq_z})$$
(16)

According to the circuit Laplacian $J(\omega)$, we can write Eqs.(9)-(16) in the form below,

$$\begin{pmatrix} I_a \\ I_b \\ I_c \\ I_d \\ I_e \\ I_f \\ I_g \\ I_h \end{pmatrix} = J \begin{pmatrix} V_a \\ V_b \\ V_c \\ V_d \\ V_e \\ V_f \\ V_g \\ V_h \end{pmatrix}. \qquad (17)$$

So the circuit Laplacian matrix in the momentum space $J(\omega)$ can be expressed as

$$J(\omega) = i\omega \begin{pmatrix} -3C_1-3C_2+\frac{3}{\omega^2 L_1}+\frac{3}{\omega^2 L_2} & 0 & C_1+C_2 e^{iq_x} & C_1+C_2 e^{iq_y} & C_1+C_2 e^{-iq_z} & 0 & 0 & 0 \\ 0 & -2C_1-2C_2+\frac{2}{\omega^2 L_1}+\frac{2}{\omega^2 L_2} & -\frac{1}{\omega^2 L_1}-\frac{e^{-iq_y}}{\omega^2 L_2} & C_1+C_2 e^{iq_x} & 0 & C_1+C_2 e^{-iq_z} & 0 & 0 \\ C_1+C_2 e^{-iq_x} & -\frac{1}{\omega^2 L_1}-\frac{e^{iq_y}}{\omega^2 L_2} & -2C_1-2C_2+\frac{2}{\omega^2 L_1}+\frac{2}{\omega^2 L_2} & 0 & 0 & 0 & C_1+C_2 e^{-iq_z} & 0 \\ C_1+C_2 e^{-iq_y} & C_1+C_2 e^{-iq_x} & 0 & -3C_1-3C_2+\frac{3}{\omega^2 L_1}+\frac{3}{\omega^2 L_2} & 0 & 0 & 0 & C_1+C_2 e^{-iq_z} \\ C_1+C_2 e^{iq_z} & 0 & 0 & 0 & -2C_1-2C_2+\frac{2}{\omega^2 L_1}+\frac{2}{\omega^2 L_2} & 0 & -\frac{1}{\omega^2 L_1}-\frac{e^{iq_y}}{\omega^2 L_2} & -\frac{1}{\omega^2 L_1}-\frac{e^{iq_x}}{\omega^2 L_2} \\ 0 & C_1+C_2 e^{iq_z} & 0 & 0 & 0 & -2C_1-2C_2+\frac{2}{\omega^2 L_1}+\frac{2}{\omega^2 L_2} & C_1+C_2 e^{iq_x} & -\frac{1}{\omega^2 L_1}-\frac{e^{-iq_x}}{\omega^2 L_2} \\ 0 & 0 & C_1+C_2 e^{iq_z} & 0 & -\frac{1}{\omega^2 L_1}-\frac{e^{-iq_y}}{\omega^2 L_2} & C_1+C_2 e^{iq_y} & -2C_1-2C_2+\frac{2}{\omega^2 L_1}+\frac{2}{\omega^2 L_2} & 0 \\ 0 & 0 & 0 & C_1+C_2 e^{iq_z} & -\frac{1}{\omega^2 L_1}-\frac{e^{-iq_x}}{\omega^2 L_2} & -\frac{1}{\omega^2 L_1}-\frac{e^{iq_x}}{\omega^2 L_2} & 0 & -2C_1-2C_2+\frac{2}{\omega^2 L_1}+\frac{2}{\omega^2 L_2} \end{pmatrix}$$

, (18)

where $\omega = \omega_0 = (L_1 C_1)^{-1/2}$. We can use Pauli matrices to express $J_\lambda(\omega_0, q)$ as Eq.(2) in the text of the paper.

## S2. The grounding parts at each site of the circuit

In this section, we provide the detailed description of the grounding parts at each site of the circuit. Due to the fact that the inductivities and capacitances used in the circuit Laplacian possessing opposite signs (Eq.1), we should choose suitable grounding parts to make the circuit Laplacian matrix become consistent with the quantum Hamiltonian. In this case, the different values of grounded capacitors/inductances should be used to make the onsite potential on each lattice site become zero at the working frequency. Therefore, we let contribution coming from all inductivities at a given site be cancelled by connecting this site to the ground with a capacitor and vice versa. This scheme can make sure the circuit system works at a fixed frequency, and the diagonal elements of circuit Laplacian matrix could be the same 0 in the resonance frequency.

In our experiment, as shown in Fig.S2 (see S3 in Supplementary Information), we divide the whole cube into adjacent six sides, which causes two types of sides A and B according to different groundings. Thus, the panels A and B in Fig.S1 show the resulting different grounding patterns.

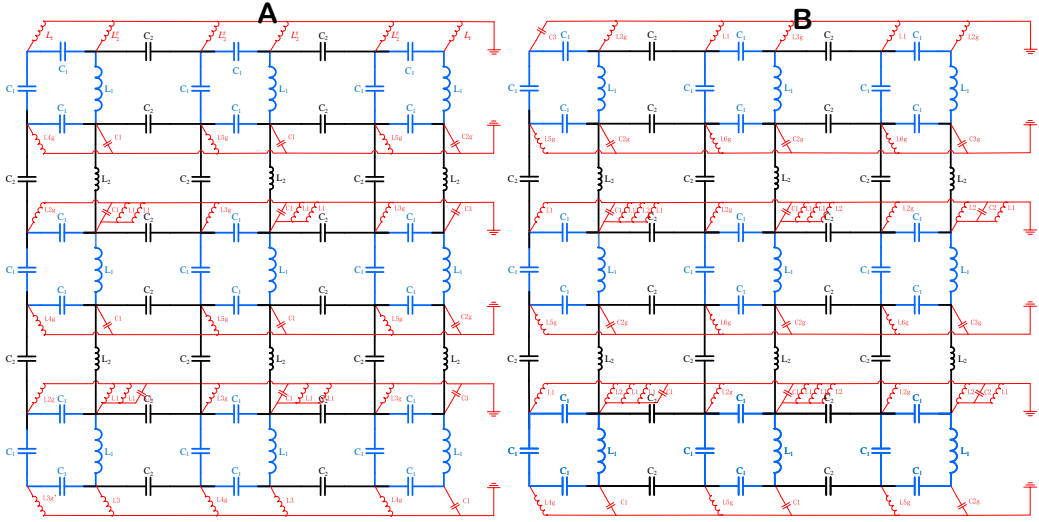

**Fig.S1.** Grounding parts at each site of the circuit for two types of sides A and B. $C_3 = C_2 - C_1$, $C_2^g = C_1 + C_2$, $C_3^g = C_1 + 2C_2$, $L_3 = 1/[\omega_0^2(C_2 - C_1)]$, $L_2^g = 1/[\omega_0^2(C_1 + C_2)]$, $L_3^g = 1/[\omega_0^2(C_1 + 2C_2)]$, $L_3^{g'} = 1/(3\omega_0^2 C_1)$, $L_4^g = 1/[\omega_0^2(3C_1 + C_2)]$, $L_5^g = 1/[\omega_0^2(3C_1 + 2C_2)]$, $L_6^g = 1/[\omega_0^2(3C_1 + 3C_2)]$ .

## S3. The detail for experimental construction of topolectrical-circuit octupole insulator

In this section, we provide the detail for experimental construction of topolectrical-circuit octupole insulator. According to the theoretical design, the constructed model for the topolectrical-circuit octupole insulator in our experiment is shown in Fig.S2, which has three unit cells in each edge. In the experiment, we divide the whole cube shown in Fig.1b into six sides. Capacitors and inductors connect adjacent side on every sites. The six sides are placed on three printed circuit boards (PCBs) in order. The different connections between adjacent sides are needed for ⅰ and ⅱ in Fig.S2. The red parts (i) of connections between adjacent sides are the capacitors and inductors which connect sides on one PCB. The orange parts (ii) of connections between adjacent sides are the capacitors and inductors which connect sides on different PCBs. The exact connections are shown in Table. I and Ⅱ for ⅰ and ⅱ, respectively. Besides, each site in the cube is connected to the ground. Due to the difference of grounding, the six sides are divided into two types A and B which have been described in S2 of Supplementary Information.

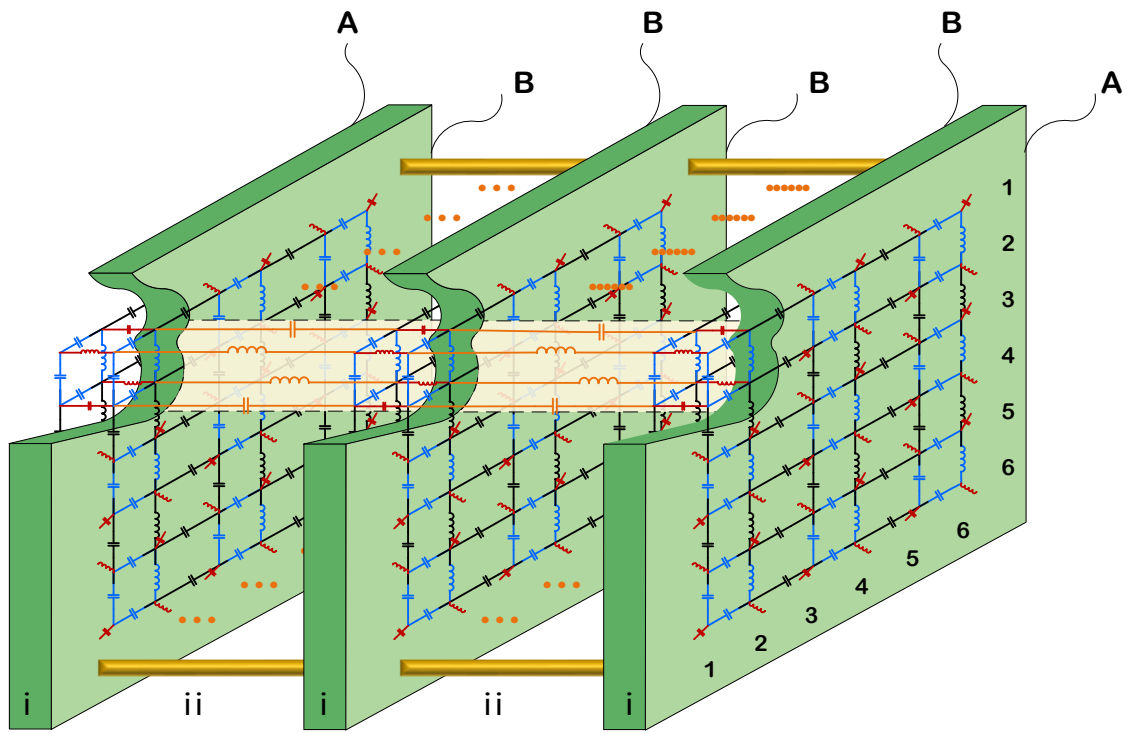

**Fig.S2.** The model pattern for experimental sample of the electrical circuit exhibiting topological corner states

**Table. I for i**

| col\row | 1 | 2 | 3 | 4 | 5 | 6 |
|---|---|---|---|---|---|---|
| 1 | L1 | C1 | L1 | C1 | L1 | C1 |
| 2 | C1 | L1 | C1 | L1 | C1 | L1 |
| 3 | L1 | C1 | L1 | C1 | L1 | C1 |
| 4 | C1 | L1 | C1 | L1 | C1 | L1 |
| 5 | L1 | C1 | L1 | C1 | L1 | C1 |
| 6 | C1 | L1 | C1 | L1 | C1 | L1 |

**Table. II for ii**

| row \ col | 1 | 2 | 3 | 4 | 5 | 6 |
|---|---|---|---|---|---|---|
| 1 | L2 | C2 | L2 | C2 | L2 | C2 |
| 2 | C2 | L2 | C2 | L2 | C2 | L2 |
| 3 | L2 | C2 | L2 | C2 | L2 | C2 |
| 4 | C2 | L2 | C2 | L2 | C2 | L2 |
| 5 | L2 | C2 | L2 | C2 | L2 | C2 |
| 6 | C2 | L2 | C2 | L2 | C2 | L2 |

The concrete values for L1, L2, C1 and C2 are taken according to the theoretical design.